\documentclass[12pt]{book}

\usepackage[dvips]{graphicx,color}
\usepackage{makeidx,tsukuba}

\makeauthorindex
\makeindex

\begin{document}

\BookTitle{\itshape The 28th International Cosmic Ray Conference}
\CopyRight{\copyright 2003 by Universal Academy Press, Inc.}
\pagenumbering{arabic}

\chapter{
SCROD: School Cosmic Ray Outreach Detector
}

\author{%
%
%
Evgueni Gouchtchine$^1$
Thomas P. McCauley,$^2$ Yuri Musienko,$^2$ 
Thomas C. Paul,$^2$ Stephen  Reucroft,$^2$ John D. Swain,$^2$
\\
{\it (1) Institute for Nuclear Research (INR),
          Russian Academy of Sciences, Moscow, Russia\\
(2) Department of Physics, Northeastern University, Boston, MA 02115} 
}

\section*{Abstract}

We report progress on applying technologies 
developed for LHC-era experiments to cosmic ray detection, using
scintillating tiles with embedded wavelength-shifting fibers and avalanche
photodiode readouts as parts of a robust, inexpensive cosmic air shower 
detector.
We are planning to deploy such detectors in high schools as part of an 
outreach
effort
able to search for long-distance correlations between airshowers.

\section{Introduction}

Ultrahigh energy cosmic rays have been detected by a number of independent experiments over 
the last few decades~[1]. The initiative we describe here, known as {\em SCROD} for {\em S}chool 
{\em C}osmic
{\em R}ay {\em O}utreach {\em D}etector~[2], 
is based on an idea which is simple but has enormous potential:
install cosmic ray detectors suitable for continuous muon counting and to
detect building-sized (or larger) extensive air showers. 
A number of similar or related initiatives have 
been proposed in the past~[3],
but SCROD offers a number of novel features which make it particularly
attractive for large-scale deployment in schools and public areas
such as libraries, community centres, {\em etc.}

Each detector is constructed from a set of inexpensive plastic 
scintillators
mounted at each site and read out using a novel technology we discuss
below.  A  PC will be used for data collection and offline data reduction.   
Cosmic air showers will be time-stamped using a GPS receiver and the data 
forwarded to a central
site which will be accessible to all the
collaborators.  We have already developed and operated a prototype 
detector and software.
SCROD will be able to study both shorter and longer-range
correlations than are accessible to many other experiments~[4]
and thus will be complementary to, and for some measurements, superior to
existing experiments.  With detectors of sufficiently low cost, 
it should be feasible to instrument
hundreds of sites, making this both an extensive educational program and
an unequaled scientific instrument.  

\section{The Detector System}

A core technology we are proposing 
is the use of Avalanche Photodiodes (APD's) 
used in the CMS experiment~[5].  
APD's are essentially photodiodes with a
large internal electric field enabling electrons
to gain enough energy within the device to free more electrons
in the semiconductor, thus providing gain. 
APD's can have high quantum efficiencies, far exceeding those of
photomultipliers. They are also mechanically robust and easy to
use requiring a supply of only a few hundred
volts, a current-limiting resistor, and a preamplifier. Furthermore
they require only a few hundred nanoamperes of current to function, making
them safe for use by non-experts.  

The hardware proposed at each detector site consists of the following
main components:
1) a set of plastic scintillating tiles with wavelength-shifting fibers to 
produce
pulses of light in response to incident cosmic rays (the optimization of
number and spacing of these tiles is still under study and is a strong 
function
of available funding);
2) APD's to read out the fibers and generate
electrical signals;
3) a GPS-based system to time-stamp signals from the APD's;
4) the World Wide Web to provide an inexpensive wide-area data acquisition
system. We concentrate on the first two components in this paper.

%
%

\begin{figure}[htbp]
  \begin{center}
    \includegraphics[height=11.5pc]{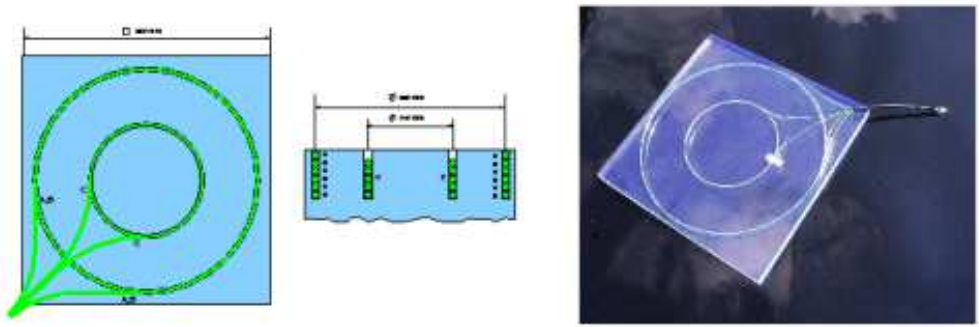}
  \end{center}
  \vspace{-0.5pc}
  \caption{Sketch and photograph of the scintillating tile with wavelength shifting 
fiber.}
\label{fig:fiberphoto}
\end{figure}

A photograph of a prototype scintillating tile with wavelength shifter
is shown in figure \ref{fig:fiberphoto}
%

\begin{figure}[htbp]
  \begin{center}
    \includegraphics[height=12.5pc]{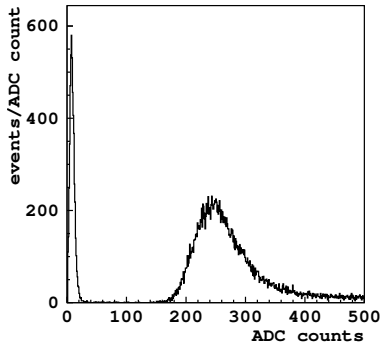}
  \end{center}
  \vspace{-0.5pc}
  \caption{ Signal due to single muons passing through
the scintillator using APD readout.}
\label{fig:pulseheights}
\end{figure}

We have already tested a prototype detector
with the APD readout
and found that it produces a
very clear signal for minimum ionizing particles (see figure 
\ref{fig:pulseheights}).

\section{Physics Potential}

Within densely-populated areas  SCROD can function
like a dramatically scaled down Pierre Auger observatory.  As many of the 
detectors will be spaced more closely together than the 
1.5~km separation used for Auger (we anticipate separations
of distances ranging from a short city block and thousands of kilometers),
it will be possible to trigger on lower
energy air showers.  In this mode, SCROD could compliment the 
results of earlier experiments. Depending on what the highest energy cosmic rays actually are, 
it is conceivable that there exist long-range correlations among
the air showers they produce. 
Either observation or non-observation of such correlations would be
a very important result which cannot readily be obtained
except by an extensive experiment of the type proposed here.
One can consider, for example, 
the Gerasimova-Zatsepin [6] effect, in which a high
energy atomic nucleus approaches the earth and dissociates on an optical 
photon from the
sun. The (two or more) nuclear fragments can then reach the earth at 
distant
locations, but close together in time. Since the composition of high 
energy 
cosmic rays is unknown, and its determination from single extensive air 
showers is
complicated by sensitivities of observables to details of the hadronic 
interaction
model chosen, this is of particularly great interest. 

Highly energetic dust grains could also dissociate and give rise
to widely-separated showers~[7].
Dramatic cosmic events may also pepper the globe with many high energy
cosmic rays all at about the same time; these have never been
looked for on the scale possible with SCROD. In addition, with the GPS 
timing
information, it will be possible
to compare and correlate data taken with SCROD with those taken at 
neutrino and
gravitational radiation detectors.

It has been suggested that cosmic rays play a significant role
in influencing climate and weather~[8], possibly due
to their effects in seeding the formation of clouds, a subject
which has only just begun experimental investigation~[9].
Modulations in the cosmic ray flux associated with solar activity can
be clearly monitored from the single count rates at each detector~[10],
and also form an important part of the collected data.

A program of educational modules will be developed through a 
close collaboration between physicists and the participating
teachers in order to be sure that our non-university collaborators
can participate in the intellectual adventure of SCROD to the
fullest. We expect to publish numerous papers on SCROD results
in professional journals, all to be signed by 
the participating students and teachers in addition to 
professional physicists.  This will 
allow students and the general public be part of a major 
international research effort, an opportunity normally 
only open to those who pursue advanced degrees in the field.
Another exciting possibility is to have joint publications with other 
experiments,
adding the SCROD data to that collected by more standard (non-outreach) 
experiments.

\section{References}

\re
1.\ L. Anchordoqui, T. Paul, S. Reucroft, J. Swain,
arXiv:hep-ph/0206072.
\re
2.\ {\tt http://www.hep.physics.neu.edu/scrod};\newline
L.A. Anchordoqui {\it et al.},
Proceedings 27th International Cosmic Ray Conferences (ICRC 2001), 
Hamburg, Germany, 7-15 Aug 2001 [hep-ex/0106002]. 
\re
3.\ See, for example, {\tt http://csr.phys.ualberta.ca/nalta/} (NALTA);\newline
{\tt http://www.physics.ubc.ca/\verb+~+waltham/alta/ } (ALTA);\newline
{\tt http://sunshine.chpc.utah.edu/ } (ASPIRE);\newline
{\tt http://www.chicos.caltech.edu/index.html} (CHICOS);\newline
{\tt http://www.unl.edu/physics/crop.html }  (CROP);\newline
{\tt http://faculty.washington.edu/~wilkes/salta/ } (SALTA);\newline
{\tt http://www.phys.washington.edu/~walta/ } (WALTA).
\re
4.\ See, for example, {\tt http://www.auger.org} (AUGER);\newline
{\tt http://www-akeno.icrr.u-tokyo.ac.jp/AGASA/} (AGASA);\newline
{\tt http://hires.physics.utah.edu/} (HIRES);\newline
{\tt http://hep.uchicago.edu/~covault/casa.html} (CASA).
\re
5. See, for example, Y. Musienko, S. Reucroft, D. Ruuska and J. Swain, 
Nucl. Instr. Meth. {\bf A} 447 (2000) 437. 
\re
6. M. Gerasimova and G. T. Zatsepin, Sov. Phys. JETP 
{\bf 11} (1960) 899; 
L. Epele, S. Mollerach, E. Roulet, 
J. High Energy Phys. {\bf 03} (1999) 017;
 G. Medina Tanco and A. A. Watson, 
Astropart. Phys. {\bf 10} (1999) 157.
\re
7. See, for example, L. A. Anchordoqui, 
Phys. Rev. D{\bf 61} (2000) 087302.
\re
8. H. Svensmark and E. Friss-Christensen, Journal of
Atmospheric and Solar-Terrestrial Physics, {\bf 59} (1997) 1225. 
\re
9. {\tt http://l3www.cern.ch/homepages/kirkby/cloud.html}
\re
10. Roger Clay, University of Adelaide, Australia -- personal
communication to J.S. based on his experience with a single small high 
school cosmic ray detector.
\endofpaper
\end{document}